\documentclass[%
reprint,
 amsmath,assume,
 aps,
]{revtex4-2}
\usepackage{lineno}
\usepackage{gensymb}
\usepackage{graphicx}
\usepackage{dcolumn}
\usepackage{bm}
\usepackage{siunitx}
\DeclareSIUnit\bar{bar}

\begin{document}


\title{Continuous relativistic high-harmonic generation from a kHz liquid-sheet plasma mirror}

\author{Antoine Cavagna$^{1}$}\email{antoine.cavagna@ensta-paris.fr}
\author{Milo Eder$^{2}$}
\author{Enam Chowdhruy$^{3}$}
\author{Andr\'e Kalouguine$^{1}$}
\author{Jaismeen Kaur$^{1}$}
\author{G\'erard Mourou$^{4}$}
\author{Stefan Haessler$^{1}$}\email{stefan.haessler@cnrs.fr}
\author{Rodrigo Lopez-Martens$^{1}$}\email{rodrigo.lopez-martens@ensta-paris.fr}

\affiliation{$^{1}$Laboratoire d'Optique Appliqu\'ee, Institut Polytechnique de Paris, ENSTA-Paris, Ecole Polytechnique, CNRS, 91120 Palaiseau, France,}
\affiliation{$^{2}$Department of Physics, The Ohio State University, Columbus, OH, 43210, USA}
\affiliation{$^{3}$Department of Materials Science and Engineering, Fontana Laboratories, The Ohio State University, Columbus, OH, 43210, USA.}
\affiliation{$^{4}$Ecole Polytechnique, Institut Polytechnique de Paris, 91128, Palaiseau, France}

\date{\today}

\begin{abstract}
	We report on continuous high-harmonic generation at 1~kHz repetition rate from a liquid-sheet plasma mirror driven by relativistic-intensity near-single-cycle light transients. Through precise control of both the surface plasma density gradient and the driving light waveform, we can produce highly stable and reproducible extreme ultraviolet spectral quasi-continua, corresponding to the generation of stable kHz-trains of isolated attosecond pulses in the time domain. This confirms the exciting potential of liquid sheet targets as one of the building blocks of future high-power attosecond lasers.
\end{abstract}

	\maketitle
	
	High-Harmonic Generation (HHG) from plasma mirrors driven by relativistic-intensity few-cycle lasers constitutes a promising way of generating intense isolated attosecond light pulses in the extreme ultraviolet (XUV) spectral region~\cite{Heissler:12,kormin:2018, Jahn:19,Boehle:20,ouille_lightwave-controlled_2024} with high (predicted) conversion efficiencies~\cite{Naumova:04,Tsakiris_2006,Mikhailova:12} (although measured efficiencies reported so far are $\sim \mathrm{10^{-4}}$~\cite{Rodel:12,Jahn:19}) and high spatio-temporal quality~\cite{Chopineau:21,Dromey:09,Leblanc:17}. Despite these credentials, the blossoming of plasma mirror-based attosecond light sources has been hampered by the almost exclusive use of moving finite-size solid targets that offer simply too low shot numbers for applications requiring high flux or high repetition rate. However, recent years have witnessed the emergence of liquid sheet targets with the required optical flatness and mechanical stability at high flow rates to drive intense light-matter interactions at multi-kHz repetition rates~\cite{morrison_mev_2018, george_high-repetition-rate_2019, Peng2024, fule_development_2024}. More recently, Kim and colleagues successfully utilized flat liquid sheet technology to demonstrate continuous HHG at 1\,kHz in the sub-relativistic regime and single shot operation in the relativistic regime~\cite{kim_liquid-shhg_2023}. This milestone experiment confirms that liquid-sheet plasma mirrors can indeed be operated continuously at high intensity while achieving the sharp surface plasma density gradients required for efficient HHG~\cite{PhysRevE.58.R5253,Kahaly:13} (although not measured in this experiment), which has remained an open question until now.

In this Letter, we report on the first measurement of the density gradient scale length at the surface of a liquid plasma mirror and on continuous HHG in the relativistic regime at kHz repetition rate using waveform-controlled near-single-cycle laser pulses. Precise control of the density gradient enables relativistic HHG at 1\,kHz with unprecedented stability. Precise control of the 1.5-cycle laser driver yields highly reproducible quasi-continuous HHG spectra~\cite{ouille_lightwave-controlled_2024}, a prerequisite for the controlled generation of isolated attosecond pulses.

\begin{figure}[t!]
\centering
\includegraphics[width=1\columnwidth]{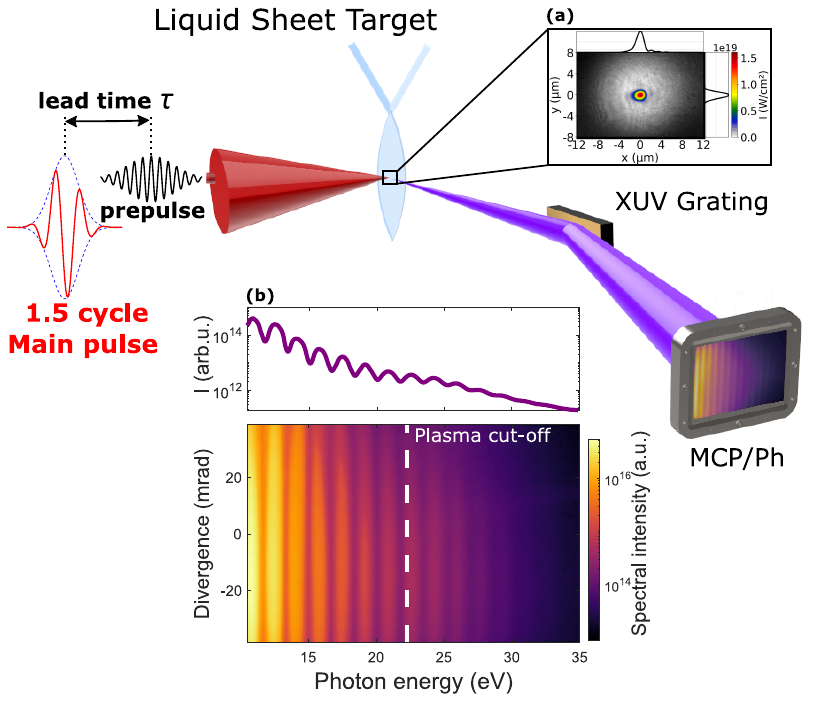}
\caption{Spatio-spectral detection setup for relativistic HHG from a liquid sheet plasma mirror. Inset (a) shows the experimentally measured focused beam profiles on target for both prepulse (gray-color scale) and main-pulse (multi-color scale). Inset (b) shows a typical angle-resolved XUV spectrogram with a vertically integrated HHG spectrum indicated on top.}
\label{fig:setup}
\end{figure} 
	
A schematic layout of our experimental setup is shown in Figure~\ref{fig:setup}. The liquid plasma mirror target was driven under vacuum by the \emph{Salle Noire} laser system delivering sub-4 fs pulses (1.5 cycle at 780\,nm central wavelength) with sub-200~mrad RMS carrier-envelope-phase (CEP) stability at 1\,kHz repetition rate and temporal contrast $>10^{10}$ @ \SI{-10}{\pico\second}~\cite{ouille_relativistic-intensity_2020}. A 4\% pick-off from the main driving pulse, henceforth called prepulse, is used to trigger spatially homogeneous plasma expansion at the liquid sheet surface before the arrival of the main pulse. Due to chromatic dispersion in its beam path, the prepulse duration on target is estimated to be $\sim$ 200\,fs. Both \textit{p}-polarized prepulse and main pulse are jointly focused onto the target at $55\degree$ with respect to the target normal using an $f/1.3$, $28\degree$ off-axis parabola (OAP) with effective focal length $f = 54.4$\,mm, yielding focal spot sizes of \SI{10}{\micro\meter} and \SI{1.8}{\micro\meter} (FWHM), respectively. With respective pulse energies of 0.1\,mJ and 2.5\,mJ reaching the target, the corresponding peak interaction intensities on target are around \SI{6e14}{\watt\per\square\centi\meter} and \SI{1e19}{\watt\per\square\centi\meter}. 	
The plasma density gradient, $L_{g} \approx L_{0} + c_{s}\tau$, is controlled by varying the lead time $\tau$ of the prepulse with respect to the arrival of the main pulse on target, where $L_{0}$ is the minimum (not measured) achievable plasma scale length due to the finite temporal contrast of the main pulse and $c_\mathrm{s}$ is the plasma expansion velocity, measured in-situ using Spatial Domain Interferometry (SDI)~\cite{Bocoum:15, Kaur:23}.

Flat liquid-sheet targets were produced in a colliding liquid jet geometry~\cite{morrison_mev_2018,george_high-repetition-rate_2019} using two 75-\unit{\micro\meter} inner-diameter capillaries oriented at a 60\unit{\degree} angle with respect to each other. Oval-shaped liquid sheets were produced roughly \SI{3}{\milli\meter}$\times$\SI{1}{\milli\meter} in size and of \SI{2.5}{\micro\meter} thickness at the center, as measured using a commercial white light interferometer (Filmetrics). These target dimensions were achieved using two syringe pumps (Teledyne ISCO), each delivering a constant flow rate of 2.4~mL/min per capillary. Ethylene glycol was selected as the medium for forming the liquid sheet due to its low vapor pressure (\(P_{\text{vap}} = \SI{1e-1}{\milli\bar}\)) and ability to sustain high-velocity turbulence-free laminar jets. For 75-\unit{\micro\meter} capillaries, the expected Reynolds number is \(\mathrm{Re} = 45\) at a flow velocity of \(8.7 \, \text{m/s}\), well below the threshold of $\mathrm{Re}\approx2000$ separating laminar from turbulent flow. A custom-designed 5-axis positioner (Smaract) allows accurate alignment of the liquid sheet target with respect to the laser focus and precise steering of the specular HHG beam into a home-made XUV imaging spectrometer downstream~\cite{Kaur:23}, in much the same way it would be done with a conventional mirror. 

The XUV spectrometer features an aberration-corrected, flat-field concave grating (Hitachi 001–0639) that lets the beam diverge in the vertical dimension while producing a spectrally dispersed image of the HHG source in the horizontal dimension on a single-stack micro-channel plate (MCP) coupled to a fast phosphor screen ($\approx300\:$ns decay time). An imaging CCD camera (PCO Pixelfly VGA) then records the HHG spatio-spectral content for photon energies ranging from 10 to 40~eV. A typically measured XUV spectrogram is shown in Figure~\ref{fig:setup}. The MCP is time-gated for 50\,ns and  triggered at 1 kHz for every laser shot, thereby suppressing most of the background plasma emission. All data points presented below were acquired by integrating the signal over 100 successive laser shots at 1\,kHz. The XUV spectrometer is shielded from stray light and open towards the HHG source through a $\approx(25\times3)\:$mm$^2$ rectangular aperture transmitting light within the angular acceptance of the grating. The resulting differential pumping keeps the background pressure around the MCP detector at $\sim10^{-5}\:$mbar, while that in the interaction chamber reaches a few $10^{-3}\:$mbar when shooting on a continuously running target at 1\,kHz. 

\begin{figure}[t!]
\centering
\includegraphics[width=0.85\columnwidth]{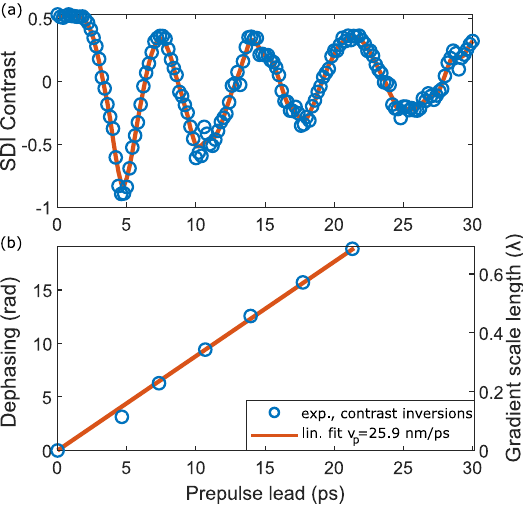}
\caption{SDI measurement of the plasma expansion velocity~\cite{Bocoum:15} at the surface of the liquid sheet target: experimental SDI contrast curve with circles/line representing raw/smoothed data (a) and computed dephasing $\Delta\phi(\tau)$ and corresponding gradient scale length $L_\mathrm{g}-L_0$ with circles/line representing extracted data/linear fit (b).}
\label{fig:SDI}
\end{figure}

Figure~\ref{fig:SDI} shows the result of the SDI measurement for the liquid plasma mirror. The linearly increasing phase of the SDI contrast oscillation yields a plasma expansion velocity of $c_\mathrm{s}~=~26\:$nm/ps through eq. 4 in ref.~\cite{Kaur:23}, with a bulk plasma density of $n_{0} = 223 n_\mathrm{c}$ corresponding to a fully ionized ethylene glycol molecule C$_{2}$H$_{6}$O$_{2}$. Note that the contrast of the final inversion from 25\,ps to 30\,ps being incomplete, it is not taken into account in the linear retrieval of the phase. This velocity is very similar, if not slightly higher, than that measured for a solid SiO$_2$-target under the same conditions~\cite{Kaur:23, ouille_lightwave-controlled_2024}. 
What remains to be seen is whether the value of $L_0$, given by the main pulse temporal contrast, also remains sufficiently small so that a clear optimum for relativistic HHG is found in a gradient scan, where HHG spectra are recorded as a function of the prepulse lead time $\tau$ before arrival of the main pulse. 
	
\begin{figure}[t]
\centering
\includegraphics[width=0.9\linewidth]{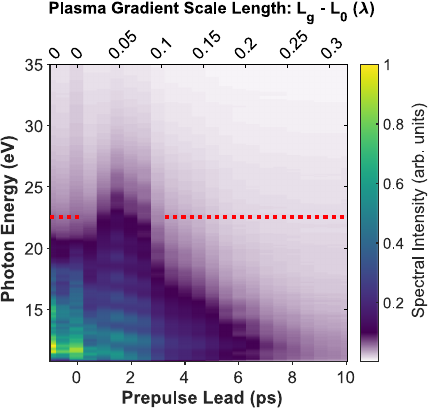}
\caption{Angularly-integrated HHG spectrum as a function of plasma density gradient scale length for 4~fs driving laser pulses with randomly varying CEP (100-shot average per spectrum). The dotted red line marks the plasma cutoff photon energy around 22.5~eV.}
\label{fig:HHGDelayScan}
\end{figure}

Figure~\ref{fig:HHGDelayScan} shows such a typical gradient scan under the aforementioned experimental conditions, with $\tau$ varied in 500\,fs time steps, using the plasma scale length obtained from the SDI measurement shown in Figure~\ref{fig:SDI}. The recorded HHG spectrograms were angularly integrated over the full detection range of $[-35, 35]\:$mrad. The spectral modulations resulting from the temporal structure of the HHG emission are known to shift with the CEP of driving pulses as short as ours~\cite{ouille_lightwave-controlled_2024, Jahn:19, kormin:2018} and thus get partly washed out by the randomly varying CEP in these measurements. Distinctive spectral features can nonetheless be identified and associated with the two well-known regimes for plasma mirror HHG: dominant coherent wake emission (CWE) at steep gradients~\cite{Quere_cwe:06,Kahaly:13}, \(L_\mathrm{g}-L_{0} < 0.025 \lambda_{0}\), and relativistic HHG at softer gradients~\cite{Rodel:12,ChopineauPRX_couplings,Haessler_UltrafastScience_2022},  \(0.025 \lambda_{0} < L_{g}-L_{0} < 0.2 \lambda_{0}\). For the steepest gradients, the temporal aperiodicity of the CWE process broadens the harmonic peaks, which in addition appear to have a dominant $2\omega_0$-periodicity in the spectral domain. This suggests a half-period temporal periodicity, consistent with a simultaneous emission of CWE and relativistic harmonics, whose different emission times during the laser cycle indeed lead to two emission events per laser period~\cite{Quere_cwe:06,Chopineau:21,ouille_lightwave-controlled_2024}. As the gradient softens, CWE vanishes and relativistic HHG gets optimized, leading to a clear $\omega_0$-periodicity of the spectral modulation with the typical narrower harmonic peaks of relativistic HHG. Figure~\ref{fig:HHGDelayScan} also clearly shows that for optimal plasma gradients, \(L_{g}-L_{0}\approx \lambda_0/20\), the HHG spectral extension far exceeds the maximum CWE cutoff at the plasma frequency~\cite{Quere_cwe:06}, which is $\approx 22.5\:$eV for a fully ionized ethylene glycol target. This structure of the gradient scan is the same as that typically observed for solid targets~\cite{Haessler_UltrafastScience_2022, ChopineauPRX_couplings, Kahaly:13}, and the observation of a clear optimum for relativistic HHG is evidence for the same fine control over the steep plasma density gradients on liquid-sheet targets as that achieved before on solid targets, even at relativistic laser intensities~\cite{kim_liquid-shhg_2023}.

\begin{figure}[t!]
\centering
\includegraphics[width = 0.9\columnwidth]{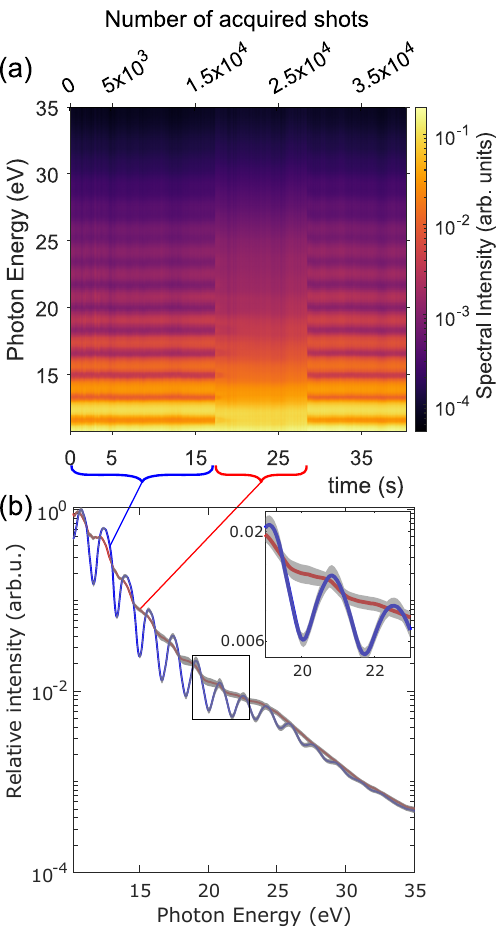}
\caption{(a): Long-term stability of relativistic HHG over $4\times10^{4}$ shots driven by 4-fs pulses with locked CEP, which was changed by $\pi$ at $18\:$s [acquired frame 237] and again at $29\:$s [acquired frame 384]. (b): Modulated (blue line) and continuous (red line) HHG spectra, obtained by averaging over the first 
18k shots before the first CEP change and over the following 
11k shots from the first to the second CEP change. The grey-shaded area behind the lines mark $\pm$1 standard deviations. Zoom-in around the plasma cutoff.}
\label{fig:stability}
\end{figure}

Having established that liquid sheet targets can essentially equivalently replace a solid target for relativistic HHG, we now move on to demonstrating a critical advancement they afford over the previously used solid-target technology through their ability to regenerate and sustain high repetition rate operation~\cite{morrison_mev_2018, george_high-repetition-rate_2019} over extended time periods. We thus fix the gradient scale length to an optimal value of $L_\mathrm{g}-L_0\approx\lambda/20$, which was found to maximize both the spectral extent and efficiency of the relativistic HHG signal, and shoot continuously at 1~kHz in order to assess its stability over time. Figure~\ref{fig:stability} shows the continuous acquisition of relativistic HHG spectra over 41 seconds while locking the laser CEP at 1~kHz repetition rate (41k shots in total). The camera acquired 13.3 frames per second with a 75~ms exposure time and a $<1\:$ms dead-time between each frame (545 frames in total). Consequently, the camera acquired at least 98.7\% of the laser shots the target was exposed to. The angularly-integrated HHG spectra displayed in Figure~\ref{fig:stability}(a) were corrected in intensity for the XUV spectrometer response (grating and MCP) and the rebinning from position (mm) to energy (eV). During this acquisition sequence, the relative CEP was changed by $\pi$, transforming the modulated relativistic HHG spectrum into to a quasi-continuum over the full detected photon energy range $(10,35)\:$eV, indicating the generation of isolated attosecond pulses~\cite{kormin:2018, Jahn:19, Boehle:20, ouille_lightwave-controlled_2024}. The CEP change was then reverted back 11~s later, recreating the exact same modulated HHG spectra as before. Figure~\ref{fig:stability}(b) shows the average spectrum and typical standard deviation for both modulated and quasi-continuum regimes. Calculating at each photon energy the coefficients of variation, that is the ratio of the standard deviation to the mean, and averaging them over the spectrum weighted by the mean, we quantify the high stability of our measured spectra, with $6\:\%$~rms and $4.5\:\%$~rms variations for the modulated and quasi-continuum regimes, respectively. Such minimal fluctuations are unprecedented~\cite{bierbach_generation_2012}, and reflect the reproducibility and consistency of attosecond pulse generation from a liquid plasma mirror in the relativistic regime, when driven by a laser with $<1\:\%$ rms pulse-energy stability.

The continuous kHz operation mode afforded by the liquid-sheet target is very different from our previous experiments with a rotating SiO$_2$ solid disk target~\cite{Borot:14}, which was used with short kHz bursts of $\sim100$ pulses, synchronized to the experimental diagnostics, so as to most economically use the available target surface~\cite{Kaur:23}. With the 0.1~mm shot spacing adapted to our few-mJ driving pulse energy, the solid target offered space for a maximum of $\approx 1.2\,$million shots, i.e. an accumulated $\approx20\:$~min at 1~kHz. In practice, pollution of the fresh target surface by debris and degrading mechanical stability with increasing rotation speed as shots got closer to the target center made target replacement necessary before maximum theoretical shot capacity. A more powerful laser driver would require a wider shot spacing, quadratically reducing the number of potential laser shots, and the debris issue would likely worsen. The available shot number on a solid target of a size that can be moved with a sub-Rayleigh-range (few \unit{\micro\meter}) stability thus rapidly becomes prohibitively low when scaling up the pulse energy of a high-repetition rate laser driver. Furthermore, the burst-mode operation limits the effective acquisition time to only $\sim1$\% of the kHz laser capacity: in practice, a day of experimentation typically resulted in about 1,000 acquisitions of 100-shot bursts ($10^5$ laser pulses). Finally, while the rotating solid target surface remained within the few-\unit{\micro\meter} Rayleigh range of the tightly focused laser, the regular translations required after completing a full circle of shots would lead to drifting interaction conditions, requiring regular realignment, which further limited the reproducibility of the HHG signal. Relativistic plasma-mirror HHG has also been demonstrated on a spooling tape target over a few 1000 shots at up to 10~Hz repetition rate, but instabilities of the tape surface lead to a 75\% rms fluctuation of the HHG signal~\cite{bierbach_long-term_2015}. Sufficiently stable and fast tape targets have since been built for laser-driven proton acceleration~\cite{xu_versatile_2023}, but available tape lengths would still limit continuous operation to a few 100~s at kHz repetition rate. 

As demonstrated in this Letter, the practical limitations of previously used solid targets are overcome with the liquid-sheet target, which enables acquiring in 2 minutes the same number of shots as a full day of experimentation with a solid target, with drastically improved stability and reproducibility. The target can be upgraded to a continuous flow operation in a closed-loop setup for an uninterrupted quasi-infinite operation time, enabling intensive usage or heavy duty measurement or alignment procedures. In the current implementation, our liquid-sheet target offers a continuous run-time of 1 hour before a refill of the syringe pumps, which only takes a few minutes. After an accumulated operation time of 7 hours, we observed very little debris or pollution on the surrounding optics and no significant drop in pulse energy delivered on target. Hitherto completely inaccessible applications thus now become feasible with plasma mirror-based HHG sources, such as quantum-optical spectroscopy based on large-number photon statistics~\cite{tsatrafyllis_high-order_2017,lamprou_quantum-optical_2021} and the challenging task of refocusing the generated attosecond pulses to high intensity for temporal characterization and nonlinear attosecond spectroscopy experiments.

	
	\begin{acknowledgments}
	This work was supported by the Agence Nationale de la Recherche (ANR-22-CE30-0005-01 BANDITO), Laserlab-Europe (654148, H2020-EU.1.4.1.2). AFOSR award (FA9550-22-1-0549). Support from ARDOP Industrie and IZEST is gratefully acknowledged.

 The authors would like to thank Hugo Marroux and S\'ebastien Brun for technical support. 

	\end{acknowledgments}

\bibliography{refs}
\end{document}